\documentclass[aps,prc,nofootinbib]{revtex4}
\usepackage{amsmath}
\usepackage{amssymb}
\begin{document}

\title{Time-delay in a multi-channel formalism} \vskip .5cm
\author{Helmut Haberzettl}
\affiliation{Center for Nuclear Studies, Department of Physics, The George
Washington University, Washington, DC 20052, USA}
\author{Ron Workman}
\affiliation{Center for Nuclear Studies, Department of
Physics, The George Washington University, Washington, DC 20052, USA}

\begin{abstract}

We reexamine the time-delay formalism of Wigner, Eisenbud and Smith, which was
developed to analyze both elastic and inelastic resonances. An error in the
paper of Smith has propagated through the literature. We correct this error and
show how the results of Eisenbud and Smith are related. We also comment on some
recent time-delay studies, based on Smith's erroneous interpretation of the
Eisenbud result.

\end{abstract}

\maketitle

 \section{The Result of Wigner and Eisenbud}

In 1948, Eisenbud~\cite{eisenbud} used a simple wave-packet approach to show
that, for an elastic resonance, the time-delay 
($\Delta t$) in a collision process was
related to the energy-derivative of the phase shift
\begin{equation}
\Delta t =\hbar \frac{d\delta}{dE}~,
 \label{eq:td_wigner}
\end{equation}
this result, apart from a factor of two~\cite{messiah}, also appeared in a
paper by Wigner~\cite{wigner}, which used causality to place a limit on
$d\delta / dE$. Less well known is the Eisenbud result for scattering into
several final states~\cite{eisenbud,note1},

\begin{equation}
\Delta t = \hbar {d\over {dE}}\left[ arg \left( S - 1\right) \right].
\end{equation}
In the case of elestic scattering, where $S = e^{2 i \delta}$, this gives the
result in Eq.~(1).

Note that $\Delta t$, in the multi-channel case, becomes a matrix. The matrix
$S$ being symmetric and unitary was diagonalized and one element was assumed to
be resonant. The resulting matrix $\Delta t$ was again found to have the form
of Eq.~(1) for each entry, however, with the phase shift 
replaced \textit{by the resonant eigenphase}. 
Unfortunately, the same notation was used for both the
phase shift and the eigenphase, and (as we will see) this may have caused some
confusion.

\section{The Result of Smith}

Smith~\cite{smith} derived the time-delay matrix based on the flux passing
through some interaction region of radius $R$. His main result was for the
average lifetime of a metastable state due to a collision beginning in the
$i^{\rm th}$ channel. The result was
\begin{equation}
Q = -i\hbar {dS\over {dE}} S^{\dagger}.
\end{equation}
Smith further claimed that his result and the result of Eisenbud could
be connected, using the following representation for Eisenbud's
result
\begin{equation}
\Delta t_{ij} = {\rm Re} \left[ -i\hbar \left( S_{ij} \right)^{-1}
              {{d S_{ij}}\over {dE}} \right]
\end{equation}
and a relation for the average over all outgoing channels
\begin{equation}
Q = \sum_j S_{ij}^* S_{ij} \Delta t_{ij}
\end{equation}
giving Eq.~(3), as claimed. The problem with this argument lies in Eq.~(4),
which is {\it not} equivalent to Eisenbud's result given in Eq.~(2) above.

\section{Do Eisenbud and Smith agree?}

To compare the results of Eisenbud and Smith, we use a representation in terms
of eigenphases. The method applied to Smith's result is given in the review
article~\cite{dalitz} of Dalitz and Moorhouse - it is reproduced here for
completeness. First use a unitary transformation to diagonalize the $S$-matrix
\begin{equation}
S = U^\dagger S_D U
\end{equation}
where $S_D$ has diagonal elements $S_{\alpha}$.
Then take the trace of $Q$
\begin{equation}
{\rm tr} Q = -i\hbar \; {\rm tr} \left[ \left( {{d U^\dagger}\over {dE}}
 S_D U + U^\dagger {{d S_D}\over {dE}} U + U^\dagger S_D
 {{d U}\over {dE}} \right) \left( U^\dagger S^*_D U \right) \right]
\end{equation}
and use the cyclic properties of the trace, the relation $U^\dagger U = 1$,
and
\begin{equation}
{\rm tr} \; \left( U {{d U^\dagger}\over {dE}} + {{d U}\over {dE}} U^\dagger
\right) = {\rm tr} \; {d\over {dE}} \left( U U^\dagger \right) = 0
\end{equation}
to find, for $S_\alpha = e^{2 i \phi_\alpha}$, where $\phi_\alpha$ is the
eigenphase,
\begin{equation}
{\rm tr}~ Q = -i\hbar \; {\rm tr} \left( {{d S_D}\over {dE}} S_D^* \right) =
        2\hbar \sum_\alpha {{d \phi_\alpha}\over {dE}}.
\end{equation}
In the simple case considered by Eisenbud, with only one resonant eigenphase
(the remaining eigenphases all approximately zero and not functions of energy),
Eq.~(9) has only one term and agrees, apart from a factor of two, with
Eisenbud's result for the time-delay in any channel.

\section{IV. A Simple two-channel Example}

These relations are clarified when applied to a simple two-channel $S$-matrix.
Consider the $T$-matrix for a two-channel problem containing only a single
Breit--Wigner resonance,
\begin{equation}
T = {{-1/2}\over {W -W_R + i\Gamma / 2}}
\left( \begin{array}{cc} \Gamma_1 & \sqrt{\Gamma_1 \Gamma_2 } \\
         \sqrt{\Gamma_1 \Gamma_2} & \Gamma_2 \end{array} \right).
\end{equation}
Construct the $S$-matrix ($S$ = 1 + 2i$T$), and diagonalize via a unitary
transformation, to find $S_D$ ($S$ = $U^\dagger S_D U$) with
\begin{equation}
S_D =
\left( \begin{array}{cc} 1 & 0 \\
         0 & {{W-W_R - i\Gamma /2}\over {W-W_R +i\Gamma/2}}
\end{array} \right),
\end{equation}
and
\begin{equation}
U = {1\over {\sqrt{\Gamma }}}
\left( \begin{array}{cc} \sqrt{\Gamma_2} & -\sqrt{\Gamma_1} \\
         \sqrt{\Gamma_1} & \sqrt{\Gamma_2}.
\end{array} \right).
\end{equation}
Here there are two eigenphases. The first, $\phi_1$, is zero and the second,
$\phi_2$, is equal to the phase of an elastic resonance. Applying the Eisenbud
relation,
\begin{equation}
\Delta t = \hbar {d\over {dE}}\left[ arg \left( U^\dagger \left(
S_D - 1\right) U \right) \right],
\end{equation}
we find
\begin{equation}
\Delta t_{ij} = \hbar {d\over {dE}} arg
{\sqrt{\Gamma_i \Gamma_j} \over {\Gamma}}
\left( e^{2i\phi_2} - 1 \right) = \hbar {{d\phi_2}\over {dE}}
\end{equation}
as given by Eisenbud and proportional to the trace of Smith's $Q$ matrix.

Several points are worth noting. Eisenbud's relation gives the same time-delay
for both channels; this time-delay is positive. The result is not altered by
energy-dependent widths, and does not depend on the branching ratio to any
particular channel.

The result obtained using Eq.~(4), wrongly attributed to Eisenbud by Smith, is
quite different. Here the time-delay depends upon the channel phase shift. In
our simple 2-channel case, the time-delay will be positive for a channel with
$\Gamma_i /\Gamma > 1/2$ and negative for $\Gamma_i /\Gamma < 1/2$.

The relation given in Eq.~(4), applied again to a simple two-channel system,
written in terms of the phase shift for a diagonal element is
\begin{equation}
\Delta t_{11} = 2\hbar {{d\delta_1} \over {dE}}~,
\end{equation}
$\delta_1$ being the phase shift for channel 1. The relation obtained using
Smith's time-delay matrix can be written in the form
\begin{equation}
Q_{11} = 2\hbar {{d\delta_1} \over {dE}} + \hbar (1 - \eta^2 )
         {{d ( \delta_2 - \delta_1 )} \over {dE}}~,
\end{equation}
where $\eta$ is the inelasticity. This clearly shows how the results differ.

\section{Time-delay versus Speed Plots}

We have for Smith's $Q$-matrix
\begin{equation}
Q_{ii} = i\hbar \left[ \sum_j S_{ij} {{dS^*_{ij}}\over {dE}} \right]_{E=E_R}~,
\end{equation}
which can be rewritten
\begin{equation}
Q_{ii} = 2\hbar \left[ {{dT^*_{ii}}\over {dE}} +
    \sum_j (2 i T_{ij} ) {{dT^*_{ij}}\over {dE}} \right]_{E=E_R}.
\end{equation}
Using relations for the multi-channel generalization of Eq.~(10),
all evaluated at $E=E_R$,
\begin{equation}
2iT_{ij} = -{4\over \Gamma} \Gamma_{ij} \;\;\;
{\rm and} \;\;\; {{dT^*_{ij}}\over {dE}} = -{4\over {\Gamma^2}} \Gamma_{ij},
\end{equation}
we have
\begin{align}
Q_{ii} &= 2\hbar \left[ -{4\over {\Gamma^2}} \Gamma_{ii} +
   \sum_j \left( {4\over \Gamma} \Gamma_{ij}\right)
       \left( {4\over \Gamma^2} \Gamma_{ij}\right) \right] \nonumber\\
     & = 2\hbar \left[ -{2\over {\Gamma^2}} \Gamma_i +
          \sum_j \left( {2\over \Gamma} \sqrt{\Gamma_i \Gamma_j}\right)
       \left( {2\over \Gamma^2} \sqrt{\Gamma_i \Gamma_j}\right) \right] \nonumber\\
     & = {{4\hbar \Gamma_i}\over {\Gamma^2}}
             \left[ \sum_j {{2\Gamma_j}\over {\Gamma}} - 1 \right]
\end{align}
where the sum in brackets gives unity. From the relation
\begin{equation}
Q_{ii} = 2\hbar \,{\rm Speed}(E_R) \qquad {\rm with} \qquad
       {\rm Speed}(E) = \vline {{dT_{ii}(E)}\over {dE}} \vline
\end{equation}
we can relate speed plot results to the time-delay relations of Eisenbud and
Smith. Note also that Eisenbud's result is given by the energy derivative of
the T-matrix phase, whereas the speed is given by the absolute value of
$dT/dE$. Combining Eqs.~(20) and (21), one may verify that we reproduce the
speed plot formula used in Ref.~\cite{lutz}.

The method of Smith has been applied in atomic physics as well. Authors from
this field have considered generalizations involving background contributions
and overlapping resonances~\cite{atomic1,atomic2}. In some cases, old results
from nuclear physics have been rediscovered~\cite{oops}.

\section{Summary and Conclusions}

In this paper we have noted an error in the work of Smith~\cite{smith} which
misrepresented the time-delay result of Eisenbud~\cite{eisenbud}. This error
has propagated through the literature, appearing also in the review article of
Dalitz and Moorhouse~\cite{dalitz}, most likely due to the fact that the result
was presented in a Ph.D. thesis, and not published.

This misquoted relation has also been used in a number of recent studies
identifying resonance signatures in elastic scattering data~\cite{kelkar}. The
results were a mixture of positive and negative time-delays, which led the
authors to erroneously conclude that speed plot and time-delay results were
inconsistent. We have shown that these problems are due to the use of phase
shifts where eigenphases are appropriate.

We have compared the time-delay and speed plot methods using an $S$-matrix
containing only a single resonance. In Ref.~\cite{mcvoy}, it was shown that the
simple extension
\begin{equation}
S_{ij} = e^{i(\varphi_i + \varphi_j )} S_{ij}^{\rm Res},
\end{equation}
with $S^{\rm Res}$ given by the multichannel generalization of Eq.~(10), and
$\varphi_i$ being a constant phase, results in eigenphases exhibiting a
correlated energy dependence in the vicinity of a resonance, a manifestation of
Wigner's no-crossing theorem~\cite{levi}. This implies a more complicated
relationship between the Eisenbud and Smith methods in the general case. The
more involved problem of two overlapping resonances, with energy independent
background, has been considered in Ref.~\cite{atomic1}.


\begin{thebibliography}{99}
\bibitem{note1}
Following the derivation in Ref.~\cite{messiah}, one might expect $(S-1)/2i$
instead of the $(S-1)$ in Eq.2. The difference is a constant overall phase that
disappears once the derivative is taken.
\end{thebibliography}
\end{document}